\documentclass[journal,10pt]{IEEEtran}
\usepackage{amsfonts}
\usepackage{mathrsfs}
\usepackage{bbm}
\usepackage{algorithm}
\usepackage{paralist}
\usepackage{algorithmicx}
\usepackage{algpseudocode}
\usepackage{multicol}
\usepackage{stfloats}
\usepackage{bm}
\usepackage{cite}
\usepackage{graphicx}
\usepackage{float}
\usepackage{subfigure}
\usepackage{stfloats}
\usepackage{amsmath}
\usepackage{amssymb}
\usepackage{color}
\usepackage{flushend,cuted}
\newtheorem{thm}{Theorem}

\newtheorem{prop}{Proposition}
\newtheorem{rem}{Remark}
\begin{document}
\newcommand{\e}[1]{\boldsymbol{#1}}

\title{\LARGE Power Minimization for Wireless Backhaul Based Ultra-Dense Cache-enabled C-RAN}

\IEEEoverridecommandlockouts

\author{Jun Xu, Jiamin Li, Pengcheng Zhu, Xiaohu You\\
National Mobile Communications Research Laboratory\\
Southeast University,Nanjing, 210096, P. R. China
}
\author{Jun~Xu,
        Pengcheng~Zhu,
        Jiamin~Li,
        and~Xiaohu~You,~\IEEEmembership{Fellow,~IEEE}\\
\thanks{This work was supported by Natural Science Foundation of Jiangsu Province (Grant No. BK20180011), and by Natural Science Foundation of China (Grant Nos. 61571120, 61871122 and 61871465).}
\thanks{J. Xu, P. Zhu, J. Li, and X. You are all with National Mobile Communications Research Laboratory, Southeast University, Nanjing, China ( emails: \{xujunseu, p.zhu, lijiamin, xhyu\}@seu.edu.cn).}
}
\maketitle


\begin{abstract}
This correspondence paper investigates joint design of small base station (SBS) clustering, multicast beamforming for access and backhaul links, as well as frequency allocation in backhaul transmission to minimize the total power consumption for wireless backhaul based ultra-dense cache-enabled cloud radio access network (C-RAN). To solve this nontrivial problem, we develop a low-complexity algorithm, which is a combination of smoothed ${\ell _0}{\text{-norm}}$ approximation and convex-concave procedure. Simulation results show that the proposed algorithm converges fast and greatly reduces the backhaul traffic.
\end{abstract}


\section{Introduction}
To meet the explosive data demand of the fifth-generation (5G) wireless system, ultra-dense (UD) networking have been widely recognized as one of the most promising technologies\cite{XGe16WC}. The significant reduction of average access distance between users and UD base stations (BSs) increases the network capacity and spectrum efficiency\cite{li2019bs}. However, it also leads to drastic interference, which constitutes the performance bottleneck for UD networking. Cloud radio access network (C-RAN) has been regarded as a promising network architecture to address this issue\cite{CPan18JSAC}, where interference mitigation can be effectively realized in the powerful baseband unit (BBU) pool by applying coordinated multi-point (CoMP). In UD C-RAN, macro base stations (MBSs) are connected to BBU with wired links and transmit data signal and control signal to small base stations (SBSs), which then provide high quality of service to users. To enable UD C-RAN, a reliable and cost-effective backhaul link (BL) connecting the MBS and SBSs is prerequisite. The expensive expenditure of wired links makes it unaffordable to be densely deployed. As an alternative, wireless backhaul (WB) emerges to present a viable solution to solve the expensive backhaul installation in UD C-RAN\cite{Gao15}.

Due to limited bandwidth and large channel fading, a critical issue in WB based UD C-RAN is to offload the peak traffic of wireless backhaul. Two enabling techniques to alleviate the backhaul are caching and SBS clustering. Motivated by the fact only a small portion of the files is highly popular with the majority of users while the rest is rarely requested, the popular files can be proactively cached in SBSs during the off-peak time, which can bring files closer to users and greatly reduce the backhaul congestion. With SBS clustering, the data of each user only needs to be delivered to its serving SBSs from the MBS rather than all the SBSs, which greatly reduces the overall backhaul load. A joint design of content-centric clustering and sparse multicast beamforming to minimize the total network cost for wired backhaul based cache-enabled C-RAN was investigated in \cite{Tao16}. Without considering cache, \cite{BHu17} jointly designed WB multicast beamforming and user-centric clustering to maximize the weighted sum rate in C-RAN.

This paper considers a joint design of content-centric SBS clustering, multicast beamforming for access links (ALs) and BLs, as well as frequency allocation in backhaul transmission to minimize the total power consumption for WB based UD cache-enabled C-RAN, which is a NP-hard problem. We first analyze the nature of the intractable constraints, and then propose a smoothed ${\ell _0}{\text{-norm}}$ approximation based convex-concave procedure (CCP) algorithm to efficiently solve the considered problem with low complexity. Simulation results show that the proposed algorithm converges fast and greatly reduces the backhaul traffic.

\section{System Descriptions}
Consider a cache-enabled UD C-RAN, which consists of a MBS with ${N_{\rm{m}}}$ antennas, $M$ SBSs with ${N_{\rm{s}}}$ antennas and $K$ single-antenna users. Let $\mathcal{M} \buildrel \Delta \over =\left\{{1,\ldots,M} \right\}$ and $\mathcal{K} \buildrel \Delta \over =\left\{{1,2,\ldots,K} \right\}$ represent the set of SBSs and users respectively. The MBS is connected to BBU through a wired link and can access a library of $F$ files with normalized size, indexed by $\mathcal{F} \buildrel \Delta \over= \left\{ {1, \ldots ,F} \right\}$. All SBSs are connected to the MBS with WB. It is assumed that the WB links between the MBS and SBSs use out-of-band spectrum, so there is no interference between ALs and BLs.
Users requesting the same file are grouped together and served by SBSs using multicast transmission. Let $\mathcal{L} \buildrel \Delta \over =\left\{{1,2,\ldots,L} \right\}$ represent the set of groups. The set of users belong to group $l$ is denoted by $\mathcal{U}_l$, and the index of group that user $k$ belongs to is denoted by $l_k$. This paper considers SBS clustering and defines a binary SBS clustering matrix ${\bf{C}} \in {\left\{ {0,1} \right\}^{M \times L}}$, where ${c_{m,l}} = 1$ indicates that SBS $m$ belongs to the serving cluster for the group $l$ and 0 otherwise. Hence, we let $\mathcal{S}_l^{\rm{A}}=\left\{m\in\mathcal{M}|c_{m,l}=1\right\}$ denote the cluster of SBSs serving group $l$.
We define a binary cache placement matrix ${\bf{S}} \in {\left\{ {0,1} \right\}^{M \times F}}$, where ${s_{m,f}} = 1$ indicates that the $f{\text{-th}}$ file is cached in the $m{\text{-th}}$ SBS and 0 otherwise. For each SBS $m\in\mathcal{S}_l^{\rm{A}}$, if the file $f_l$ requested by group $l$ has been cached in its local cache, it can directly deliver the file to all users in $\mathcal{U}_l$ via ALs. Otherwise, it needs to fetch this file from the MBS via the BLs. The cluster of un-cached SBSs belong to $\mathcal{S}_l^{\rm{A}}$ that do not cache file $f_l$ can be denoted as $\mathcal{S}_l^{\rm{B}}=\left\{m\in\mathcal{S}_l^{\rm{A}}|s_{m,f_l}=0\right\}$. Note that $\mathcal{S}_l^{\rm{B}}$ will be an empty set if all SBSs in $\mathcal{S}_l^{\rm{A}}$ have cached the file $f_l$, which means they can access the file $f_l$ directly without costing backhaul. We let $\mathcal{L}_{\rm{u}}\in \mathcal{L}$ denote the set of groups with cardinality ${L_{\rm{u}}}$ that needs backhaul transmission.

\subsection{Downlink Data Transmission Models}
The downlink data transmission consists of BL transmission and AL transmission. The details of them will be analyzed thereinafter.

\subsubsection{BL Transmission} To improve the efficiency of BL transmission, the MBS adopts multicast transmission to transmit requested files to corresponding SBS clusters. Note that one SBS may belong to multiple un-cached clusters and need to fetch multiple files from the MBS through BLs. To eliminate the interference during the decoding process at SBSs, this paper adopts a frequency division multiple access (FDMA) scheme to schedule the BL transmissions for different files, where $0\le{b_l}\le 1$ is the faction of bandawidth allocated for transmitting file $f_l$. Let $\mathbf{v}_l\in\mathbb{C}^{{N_{\rm{m}}}\times 1}$ denote the multicast beamforming vector at the MBS for group $l$. The received signal at SBS $m$ is written as
\begin{equation}
\mathbf{y}_{m}^{l}={{\mathbf{H}}_{m}^{\rm{H}}}{{\mathbf{v}}_{l}}{{x}_{l}}+{{\mathbf{z}}_{m}},\forall m\in \mathcal{M}, \forall l\in\mathcal{L},
\end{equation}
where ${{\mathbf{H}}_{m}}\in \mathbb{C}^{{N_{\rm{m}}}\times {N_{\rm{s}}}}$ denotes the channel between the MBS and SBS $m$; ${{x}_{l}}$ is the normalized signal symbol of file $f_l$; ${{\mathbf{z}}_{m}}\sim {{\mathcal{N}}_{\text{c}}}\left( 0,z_m^2 \right)$ is the noise at SBS $m$, where $\mathcal{\mathcal{N}_{\rm{c}}}(\mu,\sigma^2)$ denotes the circularly symmetric complex Gaussian distribution with mean $\mu$ and variance $\sigma^2$. Then the BL capacity $R_{m,l}^{{\rm{B}}}$ of group $l$ at SBS $m$ is given by
$R_{m,l}^{{\rm{B}}}={b_l}{{\log }_{2}}\left( 1+{{\left| \mathbf{H}_{m}^{\text{H}}{{\mathbf{v}}_{l}} \right|}^{2}}/{z_m^2} \right), \forall m\in \mathcal{M},\forall l\in\mathcal{L}$.

\subsubsection{AL Transmission} We consider the decode-and-forward scheme at SBSs. All SBSs in $\mathcal{S}_l^{\rm{A}}$ adopt multicast transmission to simultaneously forward file $f_l$ to each user in $\mathcal{U}_l$ after all un-cached SBSs in $\mathcal{S}_l^{\rm{B}}$ perfectly decoding file $f_l$. Let ${\bf{w}}_{m,l}\in\mathbb{C}^{{N_{\rm{s}}}\times 1}$ denote the beamforming vector at SBS $m$ for group $l$, then ${{\bf{w}}_{l}}{\rm{ = }}{\left[ {{\bf{w}}_{1,l}^{\rm{H}}, \cdots ,{\bf{w}}_{M,l}^{\rm{H}}} \right]^{\rm{H}}}$ is the aggregate beamformer for group $l$. The received signal at user $k$ can be written as
\begin{equation}
{{y}_{k}}=\mathbf{h}_{k}^{\text{H}}{{\mathbf{w}}_{l_k}}{{x}_{l_k}}+\sum\nolimits_{j\ne {l_k}}^{L}{\mathbf{h}_{k}^{\text{H}}{{\mathbf{w}}_{j}}{{x}_{j}}}+{{n}_{k}},\forall k\in {{\mathcal{K}}},
\end{equation}
where ${{\bf{h}}_{k}}\in\mathbb{C}^{M{N_{\rm{s}}}\times 1}$ denote the channel between all SBSs to user $k$; ${{n}_{k}}\sim {{\mathcal{N}}_{\text{c}}}\left( 0,\sigma _{k}^{2} \right)$ is the additive white Gaussian noise at user $k$. Therefore, the signal-to-interference-plus-noise ratio (SINR) of user $k$ is
${\xi}_{k}^{\text{A}}={{{\left| \mathbf{h}_{k}^{\text{H}}{{\mathbf{w}}_{l_k}} \right|}^{2}}}/\left({\sum\nolimits_{j\ne {l_k}}^{L}{{{\left| \mathbf{h}_{k}^{\text{H}}{{\mathbf{w}}_{j}} \right|}^{2}}}+\sigma _{k}^{2}}\right)$.
According to \cite{Sidiropoulos}, the common AL rate $R_{l}^{\rm{A}}$ of group $l$ is computed by
$R_{l}^{\text{A}}={{\log }_{2}}\left( 1+\min_{k\in {{\mathcal{U}}_{l}}}{\xi}_{k}^{\text{A}} \right), \forall l \in \mathcal{L}$.

\section{Problem Statement}
In this section, we first present a power consumption model. Then we formulate a joint optimization problem to minimize the total power consumption.

\subsection{Power Consumption}
 The total power consumption of a downlink cached-enabled C-RAN with WB mainly roots in transmit power at SBSs and the MBS, signal processing power at SBSs and circuit power at all BSs. Note that the signal processing at each SBS refers to the decoding process of the received signal from the MBS. If SBS $m$ has cached file $f_l$ or it is not selected to serve group $l$, it does not consume power to decode file $f_l$.  Hence, the total power consumption can be calculated as
\begin{equation}
\begin{aligned}
\label{equ_Ptotal}
P_{\rm{tot}}& {\rm{=}}{\sum\nolimits_{l,m}{{{\eta }_{m}}{{\left| {{\mathbf{w}}_{m,l}} \right|}^{2}}}}+\sum\nolimits_{l=1}^{L}{{{\eta }_{0}}{{\left| {{\mathbf{v}}_{l}} \right|}^{2}}} \\
 & {\rm{+}}{\sum\nolimits_{l,m}{{{c}_{m,l}}\left( 1-{{s}_{m,{{f}_{l}}}} \right)P_{m}^{\rm{sp}}}}+\sum\nolimits_{m=0}^{M}{P_{m}^{\rm{c}}}, \\
\end{aligned}
\end{equation}
where ${\eta }_{m}$ and ${P_{m}^{\rm{c}}}$ are respectively the inverse of the transmit amplifier efficiency and the circuit power consumption at BS $m$, where $m=0$ refers to the MBS. $P_{m}^{\text{sp}}$ denotes the signal processing power consumption used to decode at SBS $m$. Note that both $P_{m}^{\text{sp}}$ and ${P_{m}^{\rm{c}}}$ are given constants.

\subsection{Problem Formulation and Analysis}
With the given $\mathbf{S}$, this paper aims to formulate a joint design of content-centric SBS clustering, multicast beamforming vectors in ALs and BLs, as well as frequency allocation in backhaul transmission to minimize the total power consumption. All the channel state information (CSI) and user requests are assumed to be available at the BBU for joint design. The considered problem can be stated as
\begin{subequations}
\label{p0}
\begin{alignat}{2}
\mathcal{P}_0:    \min \quad& P_{\rm{tot}} &\label{p0objective}\\
\mathrm{s.t.}
\quad & {\xi}_{k}^{\text{A}} \ge  {\gamma_{l_k}}, \forall k\in \mathcal{K},&\label{p0c1}\\
& R_{l}^{\text{A}}\le R_{m,l}^{{\rm{B}}} ,\forall m\in \mathcal{S}_l^{\rm{B}},\forall l\in\mathcal{L}_{u},&\label{p0c2}\\
& \sum\nolimits_{l=1}^{L}{{{\left| {{\mathbf{w}}_{m,l}} \right|}^{2}}}\le {{P}_{m}}, \forall m \in \mathcal{M},&\label{p0c3}\\
& \sum\nolimits_{l=1}^{L}{{{\left| {{\mathbf{v}}_{l}} \right|}^{2}}}\le {{P}_0},&\label{p0c4}\\
& 0\le{b_l}\le1, \forall l\in \mathcal{L}, \sum\nolimits_{l=1}^{L}{b_l}\le 1,&\label{p0c5}\\
& c_{m,l}=\left\{0,1\right\}, \forall m \in \mathcal{M},\forall l\in \mathcal{L},&\label{p0c6}
\end{alignat}
\end{subequations}
where (\ref{p0c1}) requires that the SINR of each user should be above certain threshold; (\ref{p0c2}) denotes that the AL rate $R_{l}^{\rm{A}}$ for group $l$ should be upper bounded by the BL rate $R_{m,l}^{\text{B}}$ of SBS $m\in \mathcal{S}_l^{\rm{B}}$; (\ref{p0c3}) and (\ref{p0c4}) represent transmit power constraints at each SBS and the MBS respectively; (\ref{p0c5}) denotes the backhaul frequency allocation constraint for each group.

\begin{rem}
Different from the considered problem in \cite{Tao16} that jointly design content-centric BS clustering and multicast beamforming for cached-enabled C-RAN with wired backhaul, this paper considers a cached-enabled UD C-RAN with WB. WB induces different power consumption model, more importantly, leads to an intractable backhaul rate constraints in (\ref{p0c2}). In addition, the consideration of backhaul frequency allocation makes the problem more complex.
\end{rem}

$\mathcal{P}_0$ is a mixed integer nonlinear program (MINLP) with multiple optimization variables, which is NP-hard and can not be globally solved in polynominal complexity. In next section, we propose a low-complexity algorithm to solve $\mathcal{P}_0$.

\section{Smoothed ${\ell _0}{\text{-norm}}$ Approximation Based CCP Algorithm}
In this section, we first simplify the intractable constraints (\ref{p0c2}). Then smoothed ${\ell _0}{\text{-norm}}$ is used to approximate discrete SBS clustering with continuous beamforming vetors. At last, a CCP algorithm is proposed to solve the approximated problem.

\subsection{Equivalent Form}
To simplify (\ref{p0c2}), we first provide the following theorem, and the proof is given in Appendix A.
\begin{thm}
Assuming ${\bf{W}}^{*}=\left[ {{\mathbf{w}}_{1}^{*}},\cdots, {{\mathbf{w}}_{L}^{*}} \right]$ is the optimal solution to $\mathcal{P}_0$, we have ${\min_{k\in {{\mathcal{U}}_{l}}}{{\xi}_{k}^{\text{A}}\left({{\mathbf{W}}^{*}}\right)}}={\gamma_l},\forall l\in \mathcal{L}$ at the optimal point.
\end{thm}

Based on Theorem 1, the expression of $R_{l}^{{\rm{A}}}$ in (\ref{p0c2}) can be replaced by its threshold $r_l={{\log }_{2}}\left( 1+\gamma_l \right)$, and we have
\begin{equation}
\label{equ_rlbUR1}
r_{l}\le R_{m,l}^{{\rm{B}}},\forall m\in \mathcal{S}_l^{\rm{B}},\forall l\in\mathcal{L}_{u}.
\end{equation}
Note that $\mathcal{S}_l^{\rm{B}}$ and $\mathcal{L}_{u}$ are unavailable to BBU because of unknown SBS clustering. To tackle this difficulty, we reformulate (\ref{equ_rlbUR1}) into an equivalent form as shown in the following proposition, and the proof is omitted due to limited space.
\begin{prop}
Constraint (\ref{equ_rlbUR1}) can be equivalently transformed to the following constraint
\begin{equation}
\label{equ_rlbUR2}
{{c}_{m,l}}{\left( 1-{{s}_{m,f_l}} \right)}{r_{l}}\le R_{m,l}^{{\rm{B}}}, \forall m\in \mathcal{M},\forall l\in \mathcal{L}.
\end{equation}
\end{prop}

However, $\mathcal{P}_0$ is still hard to be solved duo to the discrete $\bf{C}$. Note that $c_{m,l}$ depends on ${\bf{w}}_{m,l}$ as $c_{m,l}={{\left\| {{\left| {{\mathbf{w}}_{m,l}} \right|}^{2}} \right\|}_{0}}$. However, the ${\ell _0}{\text{-norm}}$ is not convex. In next subsection, we approximate it with a smooth logarithmic function.

\subsection{Smoothed ${\ell _0}{\text{-norm}}$ Approximation}
According to \cite{sriperumbudur2011majorization}, we have
$c_{m,l}={{\left\| {{\left| {{\mathbf{w}}_{m,l}} \right|}^{2}} \right\|}_{0}}={\ln \left( 1+{{\left| {{\mathbf{w}}_{m,l}} \right|}^{2}}{{\varsigma }^{-1}} \right)}/{\ln \left( 1+{{\varsigma }^{-1}} \right)}$, where $\varsigma$ is a sufficiently small constant.
Using the right-hand side to replace $c_{m,l}$ in the objective and (\ref{equ_rlbUR2}), we obtain the following problem
\begin{subequations}
\begin{alignat}{2}
\mathcal{P}_1:    \min\quad & P_{\rm{tx}}+ \sum\nolimits_{l,m}{\nu_{m,l}}{\ln \left( 1+{{\left| {{\mathbf{w}}_{m,l}} \right|}^{2}}{{\varsigma }^{-1}} \right)}&\label{p1objective}\\
\mathrm{s.t.}
\quad & (\ref{p0c1}),(\ref{p0c3})-(\ref{p0c5}),&\label{p1c1}\\
& {\ln \left( 1+{{\left| {{\mathbf{w}}_{m,l}} \right|}^{2}}{{\varsigma }^{-1}} \right)}{\zeta_{m,l}}\le { R_{m,l}^{{\rm{B}}}},\forall m,l, &\label{p1c2}
\end{alignat}
\end{subequations}
where the last constant term in (\ref{equ_Ptotal}) is omitted, and $P_{\rm{tx}}$ denotes the sum of first two terms in (\ref{equ_Ptotal}). In addition, ${{\nu}_{m,l}}={\left( 1-{{s}_{m,{{f}_{l}}}} \right)}{P_{m}^{\rm{sp}}}/{\ln \left( 1+{{\varsigma }^{-1}} \right)}$ and $\zeta_{m,l}={{\left( 1-{{s}_{m,f_l}} \right)}{r_{l}}}/{\ln \left( 1+{{\varsigma }^{-1}} \right)}$.
However, the nonconvex objective (\ref{p1objective}) and constraints (\ref{p0c1}) and (\ref{p1c2}) make it still NP-hard to find the global optimum to $\mathcal{P}_1$. In next subsection, we convert $\mathcal{P}_1$ into a form of difference of convex (DC) program, and then find a suboptimal solution to it by CCP.

\subsection{CCP Algorithm to Solve $\mathcal{P}_1$}
Because ${\ln \left( 1+{{\left| {{\mathbf{w}}_{m,l}} \right|}^{2}}{{\varsigma }^{-1}} \right)}$ is concave in ${{\left| {{\mathbf{w}}_{m,l}} \right|}^{2}}$, it is upper bounded by its first-order expansion as
${\ln \left( 1+{{\left| {{\mathbf{w}}_{m,l}} \right|}^{2}}{{\varsigma }^{-1}} \right)} \le {\theta_{m,l}^{(n)}}{{\left| {{\mathbf{w}}_{m,l}} \right|}^{2}}+q_{m,l}^{(n)}$,
where ${\theta_{m,l}^{(n)}}={1}/\left({{{\left| {{\mathbf{w}}_{m,l}^{(n)}} \right|}^{2}}+\varsigma}\right)$ with the solution ${{\mathbf{w}}_{m,l}^{(n)}}$ at $n{\text{-th}}$ iteration and $q_{m,l}^{(n)}={\ln \left( 1+{{\left| {{\mathbf{w}}_{m,l}^{(n)}} \right|}^{2}}{{\varsigma }^{-1}} \right)}-{\theta_{m,l}^{(n)}}{{\left| {{\mathbf{w}}_{m,l}^{(n)}} \right|}^{2}}$. Using this upper bound, $\mathcal{P}_1$ can be approximated as
\begin{subequations}
\begin{alignat}{2}
\mathcal{P}_2:    \min\quad &{\sum\nolimits_{l,m}{{{\pi}_{m,l}^{(n)}}{{\left| {{\mathbf{w}}_{m,l}} \right|}^{2}}}}+\sum\nolimits_{l=1}^{L}{{{\eta }_{0}}{{\left| {{\mathbf{v}}_{l}} \right|}^{2}}} &\label{p2objective}\\
\mathrm{s.t.}
\quad & (\ref{p0c1}),(\ref{p0c3})-(\ref{p0c5}),&\label{p2c1}\\
& \zeta_{m,l}\left({\theta_{m,l}^{(n)}}{{\left| {{\mathbf{w}}_{m,l}} \right|}^{2}}{\rm{+}}{q_{m,l}^{(n)}}\right)\le { R_{m,l}^{{\rm{B}}}},\forall m,l, &\label{p2c2}
\end{alignat}
\end{subequations}
where ${{\pi}_{m,l}^{(n)}}={\eta_m}+{\nu_{m,l}}{\theta_{m,l}^{(n)}}$ and the constant term in the objective is omitted. At this time, the objective is convex and we turn to nonconvex constraints (\ref{p0c1}) and (\ref{p2c2}).
By introducing auxiliary variables $\psi_{m,l}$ to replace the term ${{\left| \mathbf{H}_{m}^{\text{H}}{{\mathbf{v}}_{l}} \right|}^{2}}/{z_m^2}$, (\ref{p2c2}) can be equivalently replaced by the following two constraints
\begin{subequations}
\begin{align}
&\zeta_{m,l}\left({\theta_{m,l}^{(n)}}\frac{{\left| {{\mathbf{w}}_{m,l}} \right|}^{2}}{b_l}{\rm{+}}\frac{q_{m,l}^{(n)}}{b_l}\right)\le {{\log }_{2}}\left( 1+ \psi_{m,l}\right),\forall m,l,\label{equ_c1}\\
&\psi_{m,l}-{{\left| \mathbf{H}_{m}^{\text{H}}{{\mathbf{v}}_{l}} \right|}^{2}}/{z_m^2}\le 0,\forall m,l.\label{equ_c2}
\end{align}
\end{subequations}
Here, the constraint (\ref{equ_c1}) is convex, and the left-hand side (LHS) of the introduced constraint (\ref{equ_c2}) is a DC function. In addition, the nonconvex constraints (\ref{p0c1}) can be rewritten as
\begin{equation}
\label{equ_rate1}
\left( \sum\nolimits_{j\ne l}^{L}{{{\left| \mathbf{h}_{k}^{\text{H}}{{\mathbf{w}}_{j}} \right|}^{2}}{\rm{+}}\sigma _{k}^{2}} \right){\gamma_l}{\rm{-}}{{\left| \mathbf{h}_{k}^{\text{H}}{{\mathbf{w}}_{l}} \right|}^{2}}\le 0,\forall k\in \mathcal{U}_l,\forall l\in \mathcal{L},
\end{equation}
where the LHS is also a DC function. Hence, we can apply the CCP algorithm to obtain a suboptimal solution. The subproblem
in each iteration is a convex quadratically constrained quadratic program (QCQP) and can be efficiently solved by CVX. The details are omitted.

\begin{figure*}[!t]
\centering
\begin{minipage}[b]{0.32\textwidth}
\centering
\includegraphics[width=2.5in]{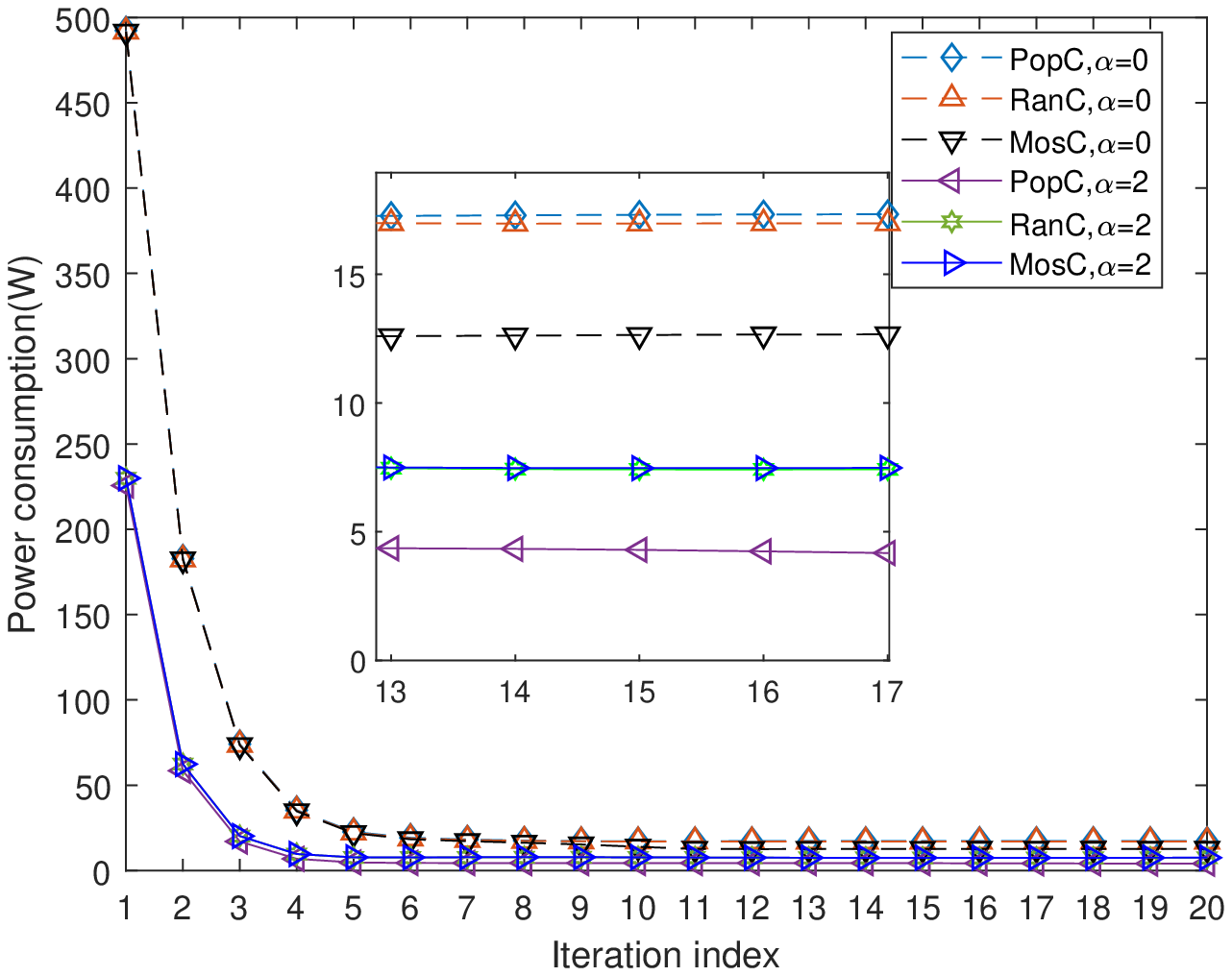}
\caption{Convergence behavior of the proposed algorithm}
\label{fig_conv}
\end{minipage}
\begin{minipage}[b]{0.32\textwidth}
\centering
\includegraphics[width=2.5in]{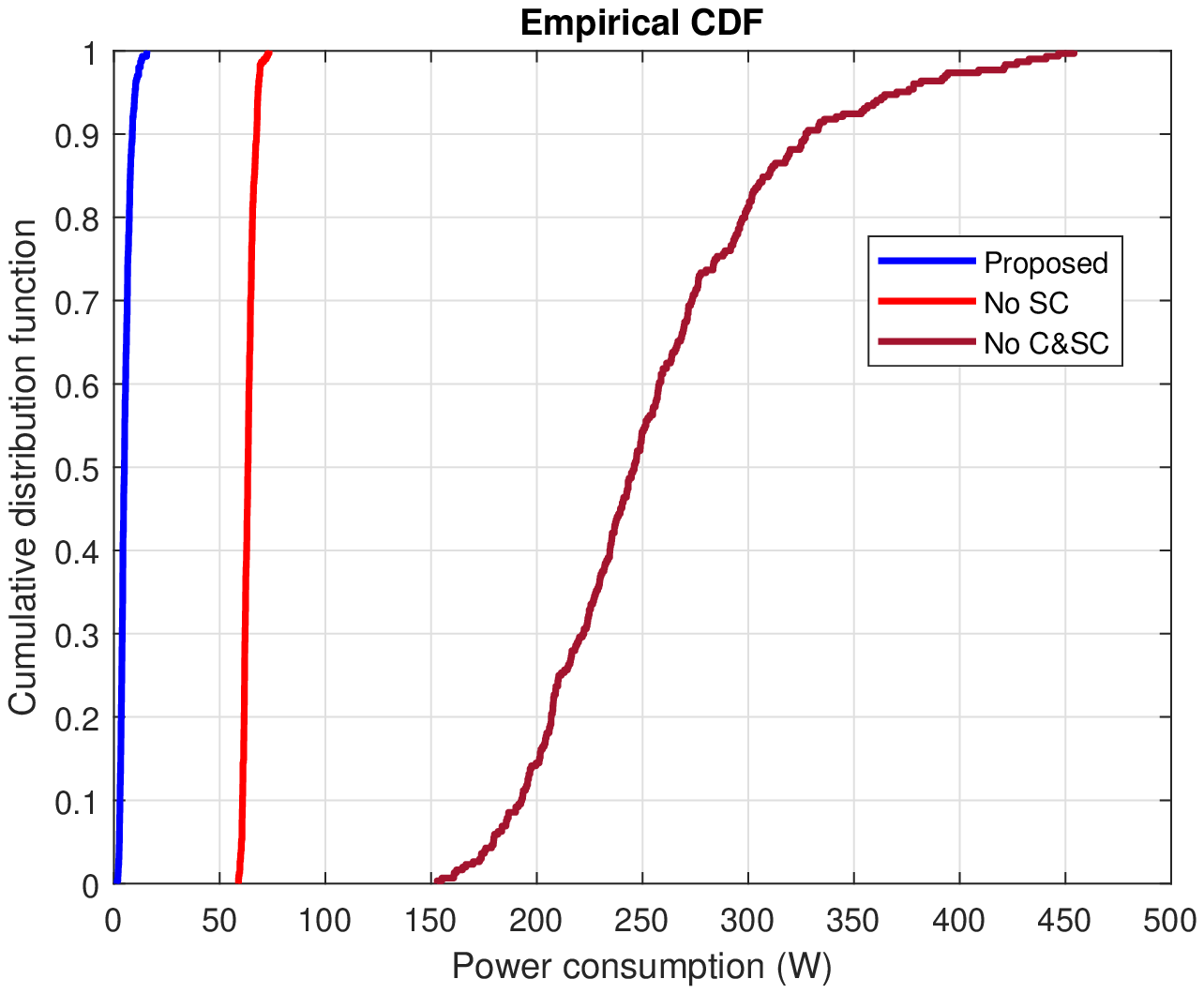}
\caption{Performance comparison between the proposed algorithm and other methods}
\label{fig_cdf}
\end{minipage}
\begin{minipage}[b]{0.32\textwidth}
\centering
\includegraphics[width=2.5in]{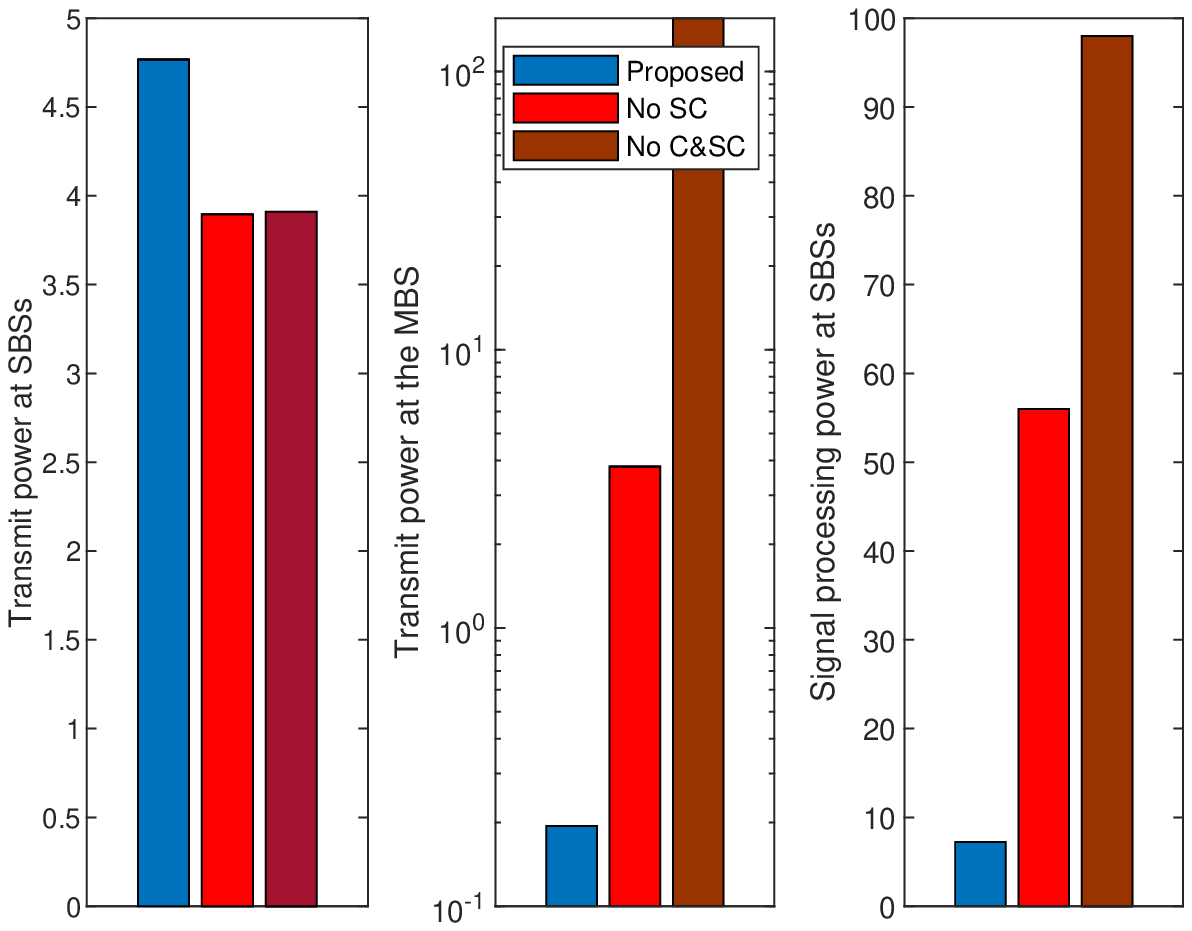}
\caption{Power consumption comparison of each term in $P_{\rm{tot}}$}
\label{fig_pcc}
\end{minipage}
\end{figure*}

\section{Simulation Results}
We consider a C-RAN network deployed in a circle area with radius $e=250m$, such as stadiums, where the MBS with $N_{\rm{m}}=50$ antennas is located at the center of this cell. There are $M=14$ SBSs with $N_{\rm{s}}=2$ antennas and $K=8$ single-antenna users randomly distributed in the cell. The densities of SBS in the considered scenario comply with the 5G UD network, where the density of 5G BS is highly anticipated to come up to 40-50 ${\rm{BS}}/{\rm{km}}^2$\cite{XGe16WC}. The channels of both BLs and ALs are modeled as ${\bf{h}}=\sqrt{d^{-\beta}{\chi}}{\bf{g}}$, where $d$ is the distance; $\beta$ is the path loss exponent; $\chi$ is a log-normal shadow fading coefficient; ${\bf{g}}$ denotes the small scale fading following Raleigh distribution $\mathcal{CN}\left(0,\bf{I}\right)$. Here, the path loss exponents for BLs and ALs are 3 and 3.2, and the variances of shadow fading for them are 3dB and 4dB. The noise power at SBSs and users are -90dBW and -65dBW. In addition, we set $P_m=10W,\forall m\in\mathcal{M}$, $P_0=50W$, $P_{\rm{sp}}=1W$ and $r_l=1.5$. We assume that there are ${F}=100$ files and each SBS can cache at most $Z=5$ files. The user requests follow a Zipf distribution with shewness parameter $\alpha$. We consider three heuristic caching strategies: PopC\cite{Tao16}: each SBS caches the top $Z$ popular files; RanC\cite{Tao16}: each SBS randomly caches $Z$ files; MosC\cite{SPark16}: each SBS $m$ caches files $\left[1,2,\cdots,Z\right]+\left(m-1\right)Z$ to cache the most files.

In Fig.\ref{fig_conv}, we show the convergence of the proposed algorithm with three cache strategies. It can be seen clearly that the proposed algorithm converge within less than 10 iterations for all considered cases. Fig.\ref{fig_cdf} shows the cumulative distribution functions (CDFs) of power consumption for three methods under PopC strategy with $\alpha=1$, where ``No SC" does not consider SBS clustering, and ``No C\&SC'' does not consider caching and SBS clustering simultaneously. They can be achieved by slightly modifying Algorithm 1 with setting $c_{m,l}=1,\forall m,l$ and $c_{m,l}=1,\forall m,l; Z=0$ respectively. The CDF curves are from 1000 independent realizations of channels for a random user request profile $[1,3,4,64,26,100,55,3]$, where the $k{\text{-th}}$ element denotes the index of the file requested by user $k$. Note that we set $P_0{\rm{=}}200W$ for ``No C\&SC'' to make it feasible. The comparative results show that considering caching and SBS clustering greatly reduces the power consumption. Furthermore, the second subfigure in Fig.\ref{fig_pcc} demonstrates that backhual traffic can be greatly reduced by considering caching and SBS clustering.

\section{Conclusion}
In this correspondence paper, we proposed a low-complexity algorithm to jointly optimize SBS clustering, multicast beamforming for ALs and BLs, as well as frequency allocation in backhaul transmission for WB based UD C-RAN. Simulation results show that the proposed algorithm converges fast and greatly alleviates the backhaul traffic.

\appendix
\appendices
\subsection{Proof of Theorem 1}
Theorem 1 can be proved by using contradiction. Specifically, we assume that one of the following two cases occurs:
\begin{enumerate}
  \item The minimum SINR of each multicast group is strictly larger than its threshold, i.e., ${\min_{k\in {{\mathcal{U}}_{l}}}{{\xi}_{k}^{\text{A}}\left({{\mathbf{W}}^{*}}\right)}}>\gamma_l, \forall l\in \mathcal{L}$;
  \item The minimum SINR of several multicast groups are strictly larger than their thresholds, while those of the rest are equal to their thresholds.
\end{enumerate}
First, we prove that the first case cannot happen. Supposing it happens, we can find another solution
$\widehat{\bf{W}}={\sqrt{a}}{{\mathbf{W}}^{*}}$, where 
$a=\max_{k\in\mathcal{U}_l,l\in\mathcal{L}}{\sigma _{k}^{2}}/\left({{{{\left| \mathbf{h}_{k}^{\text{H}}\mathbf{w}_{l}^{*} \right|}^{2}}}/{\gamma_l}-\sum\nolimits_{j\ne l}^{L}{{{\left| \mathbf{h}_{k}^{\text{H}}\mathbf{w}_{j}^{*} \right|}^{2}}}}\right)$.
Because $\min_{k\in {{\mathcal{U}}_{l}}}{{\xi}_{k}^{\text{A}}\left({{\mathbf{W}}^{*}}\right)}>\gamma_l, \forall l\in \mathcal{L}$, we have $a<1$. Hence, we have ${{\xi}_{k}^{\text{A}}\left({{\mathbf{W}}^{*}}\right)}>{{\xi}_{k}^{\text{A}}}\left(\widehat{\bf{W}}\right)\ge {\gamma_l}, \forall k\in \mathcal{U}_l,\forall l\in \mathcal{L}$.
Furthermore, it can be easily verified that $\widehat{\bf{W}}$ must be a feasible solution to $\mathcal{P}_0$ because $a<1$. For the same reason, $\widehat{\bf{W}}$ yields a lower objective value compared to that of $\bf{W}^{*}$, which contradicts the assumption that $\bf{W}^{*}$ is the optimal solution. Thus the first case cannot happen.

For the second case, we divide $L$ multicast groups into two categories, i.e., $\mathcal{L}_{1}=\left\{ l|{\min_{k\in {{\mathcal{U}}_{l}}}{{\xi}_{k}^{\text{A}}\left({{\mathbf{W}}^{*}}\right)}}>\gamma_l, \forall l\in\mathcal{L} \right\}$ and $\mathcal{L}_{2}=\left\{ l|{\min_{k\in {{\mathcal{U}}_{l}}}{{\xi}_{k}^{\text{A}}\left({{\mathbf{W}}^{*}}\right)}}=\gamma_l, \forall l\in\mathcal{L} \right\}$. Similar to the first case, we can also find another constant $a$ satisfy
$a=\max_{k\in\mathcal{U}_l,l\in\mathcal{L}_{1}}{\sigma _{k}^{2}}/\left({{{{\left| \mathbf{h}_{k}^{\text{H}}\mathbf{w}_{l}^{*} \right|}^{2}}}/{\gamma_l}-\sum\nolimits_{j\ne l}^{L}{{{\left| \mathbf{h}_{k}^{\text{H}}\mathbf{w}_{j}^{*} \right|}^{2}}}}\right)$. Defining $\widehat{\bf{W}}=\left[{\sqrt{a}}{{\bf{w}}_l^{*}},\forall l\in\mathcal{L}_{1},{{\bf{w}}_l^{*}},\forall l\in\mathcal{L}_{2}\right]$, we have ${{\xi}_{k}^{\text{A}}\left({{\mathbf{W}}^{*}}\right)}>{{\xi}_{k}^{\text{A}}}\left(\widehat{\bf{W}}\right)\ge {\gamma_l}, \forall k\in \mathcal{U}_l,\forall l\in \mathcal{L}_{1}$ and ${{\xi}_{k}^{\text{A}}}\left(\widehat{\bf{W}}\right)>{{\xi}_{k}^{\text{A}}\left({{\mathbf{W}}^{*}}\right)}\ge{\gamma_l}, \forall k\in \mathcal{U}_l,\forall l\in \mathcal{L}_{2}$. Because the minimum SINR of group $l\in\mathcal{L}_{2}$ increase, constraints (\ref{p0c2})  may not hold. To resolve this problem, we iteratively update group sets $\mathcal{L}_1$ and $\mathcal{L}_2$. In each iteration, the new constructed $\widehat{\bf{W}}$ yields a lower objective value. Because the objective value is lower bounded, the convergence of this procedure is guaranteed. When the procedure converges, the minimum SINRs of all multicast groups are equal to their thresholds, and the final $\widehat{\bf{W}}$ is a feasible solution. However, the final $\widehat{\bf{W}}$ yields a lower objective value than that with ${\mathbf{W}}^{*}$, which contradicts the assumption that $\bf{W}^{*}$ is the optimal solution. Hence, the second case cannot happen either. The proof is completed.

\bibliographystyle{IEEEtran}
\bibliography{IEEEabrv,reference}

\end{document}